# Fast and interpretable classification of small X-ray diffraction datasets using data augmentation and deep neural networks

**Authors:** Felipe Oviedo[1*&], Zekun Ren[2*], Shijing Sun[1], Charles Settens[1], Zhe Liu[1], Noor Titan Putri Hartono[1], Ramasamy Savitha[3], Brian L. DeCost[4], Siyu I.P. Tian[1], Giuseppe Romano[1], Aaron Gilad Kusne[4], and Tonio Buonassisi[1,2]

[1] Massachusetts Institute of Technology, Cambridge, MA 02139, USA
[2] Singapore-MIT Alliance for Research and Technology, 138602 Singapore.
[3] Institute for Infocomm Research (I[2]R), Agency for Science, Technology and Research (A*STAR), 138632, Singapore.
[4] National Institute of Standards and Technology, MS 8520, Gaithersburg, MD 20899, USA.
[&] Corresponding authors:

| | | |
|---|---|---|
| Felipe Oviedo | Zekun Ren | Tonio Buonassisi |
| 77 Massachusetts Av., Bldg. 35-135, Cambridge MA 02139 | Singapore-MIT Alliance for Research and Technology, 138602 Singapore | 77 Massachusetts Av., Bldg. 35-213, Cambridge MA 02139 |
| +1-617-642-1529 | +65-6516-8603 | +1-617-324-5130 |
| foviedo@mit.edu | zekun@smart.mit.edu | buonassisi@mit.edu |

[*] These authors contributed equally to this work

**Abstract**

X-ray diffraction (XRD) data acquisition and analysis is among the most time-consuming steps in the development cycle of novel thin-film materials. We propose a machine-learning-enabled approach to predict crystallographic dimensionality and space group from a limited number of thin-film XRD patterns. We overcome the scarce-data problem intrinsic to novel materials development by coupling a supervised machine learning approach with a model-agnostic, physics-informed data augmentation strategy using simulated data from the Inorganic Crystal Structure Database (ICSD) and experimental data. As a test case, 115 thin-film metal-halides spanning 3 dimensionalities and 7 space-groups are synthesized and classified. After testing various algorithms, we develop and implement an all convolutional neural network, with cross-validated accuracies for dimensionality and space-group classification of 93% and 89%, respectively. We propose average class activation maps, computed from a global average pooling layer, to allow high model interpretability by human experimentalists, elucidating the root-causes of misclassification. Finally, we systematically evaluate the maximum XRD pattern step size (data acquisition rate) before loss of predictive accuracy occurs, and determine it to be 0.16° $2\theta$, which enables an XRD pattern to be obtained and classified in 5.5 minutes or less.

**Keywords:** X-ray diffraction, novel materials, thin-films, machine learning, data augmentation, crystallography, interpretable machine learning

## I. Introduction

High-throughput material synthesis and rapid characterization are necessary ingredients for inverse design and accelerated material discovery [1,2]. X-ray diffraction (XRD) is a workhorse technique to determine crystallography and phase information, including lattice parameters, crystal symmetry, phase composition, density, space-group, and dimensionality [3]. This is achieved by comparing XRD patterns for a novel material to the measured or simulated XRD patterns of known materials [4]. Despite its indispensable utility, XRD is a common bottleneck in materials characterization and screening loops: one hour is typically required for thin-film XRD data acquisition for a $2\theta$ scan with high angular resolution, and another one to two hours are typically required for Rietveld refinement by an expert crystallographer, assuming the possible crystalline phases are known. It is widely recognized that machine learning methods have potential to accelerate this process; however, practical implementations have thus far focused on well-established materials [5–7], require combinatorial datasets spanning among various phases [8,9], or require large datasets [5,10], while material screening using the inverse design paradigm often involves less-studied materials, spanning multiple classes of with different material / phase compositions, and smaller data sets prototypes.

Typically, experimental XRD pattern data is analyzed by obtaining descriptors such as peak shape, height and position. Matching descriptors of the test pattern to known XRD patterns in crystalline databases allows the identification of the compound of interest[4]. Refinement methods such as Rietveld refinement and Pawley refinement have been used for decades to analyze experimental XRD patterns[4]. For novel compounds in thin-film form, however, the use of Rietveld refinement is limited due to the lack of reference patterns in the database as well as unknown film textures. The direct-space method, statistical methods, and the growth of single crystals have been used to obtain crystal symmetry information for novel materials [7,11–14], but the significant iteration time, feature engineering, human expertise, and knowledge of specific material required makes these methods impractical for high-throughput experimentation, where sample characterization rates are of the order of one material per minute or faster [2,15], explored over various material families.

An alternative approach consists in using machine learning methods to obtain more robust spectral descriptors and quickly classify crystalline structure based on the peak location and shape in the XRD pattern. Breakthrough methods have been developed for the similar problem of phase attribution in combinatorial alloys[16,17], but only few studies have been developed for solution-processed material screening, such as perovskite screening, where phase attribution is usually not as important as correct classification of materials into groups according to crystal parameters. The most successful methods [10,18] for material screening use convolutional neural networks (CNN) trained with hundreds of thousands of XRD powder patterns simulated with data from the Inorganic Crystalline Structure Database (ICSD). Further CNN and other deep learning algorithms have been employed to obtain crystalline information for other kinds of diffraction data [19–21]. In a couple studies, noise-based data augmentation, a common technique of image preprocessing for machine learning, has been used to avoid overfitting in a broader kind of X-ray characterization problems [20,22,23] and more broadly in other fields such as TEM imaging [21], however the augmentation procedure has not been based in physical knowledge of actual experimental samples. Furthermore, the machine learning methods developed up to date for XRD analysis do not allow any kind of interpretation by the experimentalists[10], hindering improvements of experimental design.

While similar approaches produce good results for crystal structure classification, we have found that applying them to high-throughput characterization of novel solution-processed compounds is generally not practical, given the limited access to large datasets of clean, pre-processed, relevant, XRD spectra. Furthermore, most materials of interest developed in high-throughput synthesis loops are thin-film materials. The preferred orientation of the crystalline planes in thin-films causes their experimental XRD patterns to differ from the thousands of simulated XRD powder patterns available in most databases [24,25]. Thin-film compounds usually will present spectrum shifting and periodic scaling of peaks in preferred orientations, reducing the accuracy of machine learning models trained with powder data [8,9,26], even in the cases when noise-based data augmentation techniques are used. [18]

Considering these challenges, we propose a supervised machine learning framework for rapid crystal structure identification of novel materials from thin-film XRD measurements. For this work, we created a library of 164 XRD patterns of thin-film halide materials extracted from the >100,000 compounds available in the ICSD [27]; these 164 XRD patterns include lead-halide perovskite[28,29] and lead-free perovskite-inspired materials [30]. These XRD patterns were manually classified among different crystal dimensionalities using ICSD information. Based on this small dataset of relevant XRD powder patterns extracted from the ICSD and an additional 115 experimental XRD patterns, we propose a model-agnostic, physics-informed data augmentation to generate a suitable and robust training dataset for thin-film materials, and subsequently test the space-group and dimensionality classification accuracy of multiple machine learning algorithms. A one-dimensional implementation of an "*all convolutional neural network*"[31] is proposed, implemented and identified as the most accurate and interpretable classifier for this problem. We propose a way to use class activation maps[32], computed from the weight distribution of a global average pooling layer and adapted to the context of our problem, to provide interpretability of classification success or failure pattern for the experimentalist. Subsequently, the effect of the augmented dataset size and the XRD pattern granularity is investigated. Our proposed methodology could be applied to other crystal descriptors of thin-film materials, such as lattice parameter or atomic coordinates, as long as labelled information is available.

Our contributions can be summarized as: a) Development of physics-informed data augmentation for thin-film XRD, which successfully addresses the sparse / scarce data problem, breaching the gap between the thousands of XRD patterns in crystalline databases and real thin-film materials, b) Development of highly interpretable, highly-accurate, all convolutional neural network for XRD material screening, c) Proposal of Average Class Activation Maps as a feasible interpretability tool in convolutional neural networks trained on spectral data.

## II. Results and Discussion

***Framework for rapid XRD classification***

The framework developed for rapid classification of XRD thin-film patterns according to crystal descriptors is shown in Figure 1a. The methodology makes use of both experimental and simulated XRD patterns to train a machine learning classification algorithm. A simulated dataset is defined by extracting crystal structure information from the ICSD. The experimental dataset consists of a set of synthesized samples, which are manually labelled for training and testing purposes. The datasets are subjected to data augmentation based on the three spectral transformations shown in Figure 2.

The crystal descriptors of interest, space-group and crystal dimensionality, are chosen because of their importance for material screening in accelerated material development. In many inorganic material systems, the crystalline dimensionality — *i.e.*, a generalization of the crystalline symmetry into 0-dimensional (0D), 1D, 2D or 3D symmetry— constitutes a figure of merit for experimental material screening as it correlates with observed charge-transport properties.[33] In perovskites and perovskites-inspired materials, for instance, 3D crystalline structures have been shown to have good carrier-transport properties for solar cells and LED applications,[33,34] while 3D-2D mixtures have been found to have greater stability in lead halide perovskites than pure-phase 3D crystals.[35] With further detail, the space-group number describes the standardized symmetry group of a configuration in space, classifying crystal symmetries into 230 groups. Identifying the space-group number of a sample provides crystal information beyond dimensionality, including atomic bonding angles and relative distances, which are believed to be of importance for predicting material properties [36]. In this specific study, the framework relies on the relation between XRD patterns and the crystal descriptors of interest. For example, among perovskite-inspired materials for photovoltaic applications, 3D cubic lead halide perovskites of multiple compositions show distinct features in the XRD pattern compared to 2D layered bismuth perovskites [37,38].

Typically, the powder XRD pattern is used to identify space-group through Rietveld refinement, but the compression of crystalline three-dimensional crystallographic information into a one-dimensional diffraction pattern causes the space-group to be impossible to determine unambiguously in certain low-symmetry phases, independently of the measurement technique [39]. In this work, the space-groups of interest are able to be determined from XRD information only.

To better account for noise measurement and the physical difference between randomly-oriented powder patterns and experimental thin-film patterns, the patterns were subjected to a process of data augmentation based on domain knowledge. Subsequently, both augmented experimental and simulated XRD pattern datasets are used for testing, training and cross-validation of machine learning algorithms.

Figure 1b shows the architecture of the final all convolutional network, which is proposed and identified as the best performing machine learning algorithm in subsequent sections.

*Experimental measurement and labelling of XRD patterns*

The experimental dataset consists of 75 XRD patterns for dimensionality classification, summarized in Table S1 in the Supplementary Information, and 88 XRD patterns for space-group classification, summarized in Table S2. A total of 115 unique labelled XRD patterns are considered among both datasets. For this work, perovskite-inspired 3D materials based on lead halide perovskites (space-group $Pm\bar{3}m$), tin halide perovskite ($I4/mcm$), cesium silver bismuth bromide double perovskite ($Fm\bar{3}m$), bismuth and antimony halide 2D ($P\bar{3}m1$, $Pc$, $P2_1/a$) and 0D ($P6_3/mmc$) perovskite-inspired materials are synthesized and used as training and testing dataset. The details of the synthesis and characterization methodology are described in great detail in our experimental study [40]. Figure S1 in the Supplementary Information shows the t-SNE representation of the XRD patterns labelled with dimensionality and space-group, providing evidence of the complexity of the classification problem.

Rietveld refinement is restricted in this case due to the unknown preferred orientation and texture of the thin-film samples. In consequence, the XRD patterns are subjected to peak indexing and the dimensionality and space-group are confirmed based on ICSD data, for those cases when reference patterns are available in the dataset.

*Data mining and simulation of XRD patterns*

The simulated training dataset consists of 164 compounds extracted from ICSD with a similar composition, expected crystal symmetry, and space-group as the synthesized materials of interest. All the possible single, double, ternary, and quaternary combinations of the elements of interest were extracted during database mining. The compositions of all the materials of interest along with the labelled dimensionality and space-group information is available in Table S3. The fundamental crystal descriptors extracted from the material database are used to simulate XRD powder patterns with random crystalline orientations, as explained in the Methods section.

*Data Augmentation based on domain knowledge*

Experimental thin-film XRD patterns vary greatly compared to simulated, idealized, randomly-oriented XRD patterns. Due to expansions and contractions in the crystalline lattice, XRD peaks shift along the 2$\theta$ axis according to the specific size and location of the different elements present in a compound, while maintaining similar periodicity based on crystal space-group[3] [8,26,41]. In addition, for thin-film samples, the XRD pattern can be shifted due to strain in the film induced during the fabrication process [42]. Polycrystalline thin-films are also known to have preferred orientations along certain crystallographic planes. The preferred orientation is influenced by the crystal growth process and substrate [43], and is common for most solution-processing and vapor-deposition fabrication methods. Ideal random powders contain multiple grains without any preferred global orientations, thus all crystallographic orientations are represented evenly in the peak intensity and periodicity of the XRD pattern. As a consequence of their preferred orientations along crystallographic planes, thin-film XRD relative peak intensities are scaled up periodically in the preferred plane orientation, and scaled down periodically or even eliminated in the non-preferred orientations.

To increase the size and robustness of the limited training dataset and to account for these fundamental differences between real thin-films and simulated XRD powder spectra, we perform a three-step data augmentation procedure based on physical domain knowledge:

1. Peak scaling, 2. Peak elimination and 3. Pattern shifting. These are transformation are described in detail in Methods. Figure 2 summarize the data augmentation steps and its effects on a representative pattern. Given random variables $S$ and $\varepsilon$ in Eqs. 1–3 in the Methods section, 2,000 patterns are augmented from the simulated dataset, and 2,000 patterns are augmented from experimentally measured spectra.

We choose to perform physics-informed data augmentation instead of explicit regularization for the following reasons: 1. Data augmentation has been found to be more robust at avoiding overfitting than explicit regularization when using neural networks[44], 2. Data augmentation is model-agnostic, allowing our approach to successfully bridge the gap between experimental XRD patterns and thousands of XRD patterns available in databases without depending on a specific model that might not generalize well in all cases (no-free-lunch theorem), and 3. Physics-informed data augmentation allows high interpretability, and is found to be more robust than traditional noise-based augmentation approaches (Table S6).

### *Classification results and All Convolutional Neural Network*

Pre-processed, augmented experimental data and augmented simulated data are fed into various supervised machine learning algorithms for training and testing purposes. The best-performing algorithm is evaluated. The XRD patterns are classified into 3 crystal dimensionalities (0D, 2D, and 3D) and 7 space-groups ($Pm\bar{3}m$, $I4/mcm$, $Fm\bar{3}m$, $P\bar{3}m1$, $Pc$, $P2_1/a$, $P6_3/mmc$).

For this purpose, we represent the XRD pattern as either a vector or a time series. For each kind of data representation, different classification algorithms are considered. Using a vector representation of the XRD pattern, the following classification methods are tested: Naïve Bayes, *k*-Nearest Neighbors, Logistic Regression, Random Forest, Decision Trees, Support Vector Machine, Gradient Boosting Decision Trees, a Fully-Connected Deep Neural Network, and an All Convolutional Neural Network with a global pooling layer[45–47]. The XRD patterns are also analyzed as a time-series with a normalized Dynamic Time Warping (DTW) distance metric[48] combined with a *k*-Nearest Neighbors classification algorithm, which was found in literature as the most adequate metric for measuring similarity among metal-alloy XRD spectra[7,26].

The problem of novel material development is inherently a multi-class classification problem, in which the classes for training and testing purposes can often be imbalanced as some material families are better characterized than others (*e.g.*, lead-based perovskites are better represented in material databases than newer lead-free perovskites)[38]. Common metrics for binary classification such as accuracy might not be the most adequate in this context [46]. The final choice for adequate metrics depends on the relative importance of false positive and false negatives in minority and majority classes, according to the goals of the experimentalist. For method development in this work, we consider the following metrics: subset accuracy, defined as the number of correctly classified patterns among all test patterns, and $F_1$ score. $F_1$ score in this problem can be interpreted as the weighted harmonic mean of precision and recall; the closer it is to 1.0, the higher the classifier's precision and recall[46]. Intuitively, precision is the ability of the classifier not to produce a false positive, whereas recall is the ability of a classifier to find all the true positives. An $F_1$ metric is calculated for each class label, and it is combined into an overall score by taking either the *micro* or the *macro* average of the individual scores. The *macro* average calculates the mean of the metrics of all the individual classes, hence treating all classes equally. The *micro* average adds the individual contribution of all samples to compute the overall metric.

When there is class imbalance, accuracy and $F_1$ *micro* score characterize the classifier's performance over all classes, whereas $F_1$ *macro* emphasizes the accuracy on infrequent classes[46]. Thus, a natural choice for high-throughput experiments across multiple material classes seems to be accuracy / $F_1$ *micro* score, except in those cases when we are especially interested in analyzing an infrequent material class, being $F_1$ *macro* a more representative metric in that case. In this work, we choose to report the classification accuracy and both $F_1$ *micro* and $F_1$ *macro,* while recall and precision results are included in the Supplementary Information (Table S4 and Table S5).

We measure the performance of the dimensionality and space-group classification methods based on three different approaches of splitting the training and testing datasets:

**Case 1:** Exclusively simulated XRD patterns are used for testing and training. 5-fold cross validation is performed.

**Case 2:** The simulated XRD patterns are used for training, and the experimental patterns for known materials are used for testing.

**Case 3:** All of the simulated data and 80% of the experimental data are used for training, and 20% of the experimental data are used for testing. 5-fold cross validation is performed.

Each one of the training/testing cases mentioned earlier are tested for crystal dimensionality and space-group prediction accuracy and *micro*/*macro* $F_1$ score. The results are reported in Table 1. In each cell, the crystal dimensionality classification metric is reported first, followed by the metric for space-group classification. Case 1, presenting 5-fold cross validation results of the simulated dataset, has the highest accuracy as it does not predict any experimental data and thus is free of experimental errors for both crystal descriptors. Case 2 performs the experimental prediction solely based on simulated patterns, thus having the lowest accuracy. Finally, Case 3 has a significant higher accuracy than Case 2 for both crystal dimensionality and space-group prediction. $F_1$ scores follow these trends as well.

In general, the model's accuracy and $F_1$ score is lower for space-group classification than that for crystal dimensionality classification. This discrepancy is caused by the lower number of per-class labelled examples for space-group classification compared to crystal dimensionality classes. Class imbalance can also systematically affect the training performance of the classifier, to avoid this issue, we performed an oversampling test with synthetic training data according to [49,50], and observed little discrepancy of accuracy between the balanced and imbalanced datasets after 5-fold cross validation.

The use of experimental data as part of the training set increases the model's accuracy and robustness. This fact can be explained by the high variability of experimental thin-film XRD patterns, even after data pre-processing. The relatively high accuracy with the relatively small number of experimental samples (on the order of 10–10$^2$) confirms the potential of our data augmentation strategy to yield high predictive accuracies even with small datasets. Table S6 in the SI compares our strategy to traditional noise-based augmentation approaches, and shows an average increase of classification accuracy of more than 12% absolute.

Naturally, the $F_1$ *macro* score is systematically lower than the $F_1$ *micro* score, reflecting the impact of misclassification of those dimensionality and space-group classes with less training examples. However, the $F_1$ *macro* score is still fairly high for most classifiers. This fact reflects the importance of adequate experimental design to achieve good generalization among classes.

For all three test cases, the all convolutional neural network (a-CNN) classifier performs better than any other classification technique. The 1-dimensional a-CNN architecture implemented is composed of three 1-dimensional convolutional layers, with 32 filters each, and strides and kernel size of 8, 5 and 3 units respectively. The activation function between layers is *ReLu*. A global average pooling layer[51] (acting as a weak regularizer) and a final dense layer with *softmax* activation is used. The loss function minimized is binary cross entropy. We use early stopping with a batch size of 128 during training, and use the *Adam* optimizer algorithm to minimize the loss function. The CNN is implemented in Keras 2.2.1 with the Tensorflow background. Figure 1b) contains a schematic of the proposed a-CNN architecture.

Our a-CNN architecture, in contrast with other convolutional neural networks, does not have a max pooling layers in between convolutional layers, and also lack of a set of dense layers in the final *softmax* classification layer. These modifications, in contrast with the architectures used in [10,18], significantly reduces the number of parameters in the neural network and allows faster and simpler training, and are less prone to overfitting. Another advantage of our implementation is the possibility to extract class activation maps using global average pooling layer. This, properly adapted to our problem context, allows us to visualize how the classified XRD patterns correspond to weights distribution in the last convolutional layer. The results are further discussed in the interpretability subsection.

The a-CNN trained after data augmentation has an accuracy of more than 93% and 89% for crystal dimensionality and space-group classifications, respectively. As far as we know, the accuracy is the higher up to date based on results found in literature for space-group classification algorithms trained with thousands of ICSD patterns and manual labelling by human experts [10,52], and is also comparable to similar approaches in other kinds of diffraction data [8,19]. The neural network seems to be the most adequate method for high-throughput synthesis and characterization loops, as it also performs relatively well in terms of algorithm speed and in conditions of class imbalance. In the future, our methodology can be extended to other materials systems, and may include other crystal descriptors as predicted outputs, such as lattice parameters and atomic coordinates.

Furthermore, the a-CNN performs better than the traditional *k*-nearest neighbors method using DTW. In our test case and dataset, the differences between thin-film and powder spectra seem not to be captured properly by DTW alone. Arguably, DTW could be more useful if a larger XRD thin-film pattern dataset is available for *k*-Nearest Neighbors classification, or if it exists greater similarity between XRD patterns of the same class, allowing the DTW warping path to be better captured within the DTW window under consideration [26]. Convolutional neural networks have been found to perform better than DTW for classification of time-series, which is consistent with our results[53].

**Effect of augmented dataset size:**

The size of the dataset is critical for obtaining a high accuracy and $F_1$ score. To explore the effect of augmented dataset size, the a-CNN accuracy was computed for various combinations of augmented experimental (*i.e.*, number of augmented XRD spectra originated from the 88 measured spectra, varying $S$ and $\varepsilon$ in Eqs. 1–3) and augmented simulated dataset sizes (*i.e.*, number of augmented XRD spectra originating from the 164 simulated ICSD spectra). Figure 3a and Figure 3b summarize this sensitivity analysis for Case 3 training/testing conditions for dimensionality and space-group classification. Twenty different five-fold cross-validation runs were performed to calculate the 1-standard-deviation error bars for each data point.

In general, as the size of the experimental and augmented datasets increase, the mean accuracy quickly approaches the asymptotic accuracy reported in Table 1. This trend reaffirms and quantifies the importance of data augmentation for the predictive accuracy of our model. The critical augmented-dataset size seems to be around 700 augmented spectra. The model's accuracy is more sensitive to the augmented experimental dataset size, likely because most of the dataset variance comes from the experimental XRD patterns. The data augmentation of the simulated dataset causes the accuracy to grow monotonically in Figure 3b; however, this trend is not satisfied in the case of Figure 3a, where no augmented simulated data seems on average to perform the best. We hypothesize that augmenting simulated data could actually introduce excessive noise to the model, hampering classification when the number of possible classes is small.

Figure 3 illustrates that if no data augmentation is used (*i.e.*, the origin, 0, 0), the predictive accuracy could be below 50% for space-group and below 70% for dimensionality. Our physics-informed data augmentation directly increase accuracy by up to 23% in the case of dimensionality and 19% in the case of space-group classification. This result reinforces the need for data augmentation for sparse datasets, as is typical with early-stage material development.

**Impact of data coarsening**

To evaluate trade-offs between ML classification accuracy and XRD acquisition speed, we investigate how data coarsening of the XRD pattern impacts the accuracy of ML algorithm prediction. In Figure 4, we report Case 3 accuracy with increasing 2𝛳 angle step size. The baseline step size of the 2𝛳 scan in our XRD patterns is 0.04°. Data coarsening is performed by removing the data with different step size and rerunning the augmentation and classification algorithms. For crystal dimensionality and space-group classification, the highest accuracies are achieved at 0.04-0.08°, while 85%+ accuracy is achieved when the 2𝛳 step-size is 0.16° or less for both cases. Using the larger step-size, the XRD pattern acquisition time can be reduced by 75%, allowing the full spectra to be measured and classified in less than 5.5 minutes with our setup.

**Interpretability using Class Activation Maps:**

Class activation maps (CAM) are representations of the weights in the last layer of a convolutional neural network, before performing classification. A CAM for a certain class and pattern indicates the main discriminative regions (in our case, peak and series of peaks in the XRD pattern) that the network uses to identify that class[32]. Details of the CAM computation are included in the Methods section.

A similar approach has been followed to interpret and improve object recognition in images and videos using two-dimensional CNNs[32]. A single pattern or image produces a unique CAM revealing the main discriminative features. In addition, in the context of our problem, we propose to generalize CAMs to all training samples within a class by averaging over each of the computed CAM weights for the all training samples within a class, as explained in the Methods section. This averaging procedure is justified as the location and periodicity of discriminative peaks within a class varies only slightly along all labelled samples in the training set, and can be extended to many spectral measurement techniques. This *average CAM* allows to visualize the main discriminative features of an XRD pattern used to classify testing data within a given class.

By comparing the CAM for a single pattern to the average CAM for given class, we can to identify the root-causes of correct and incorrect classification by the a-CNN. Figure 5 illustrates this procedure for XRD patterns of space-group classes Class 2 ($P2_1/a$) and Class 6 ($Pm\bar{3}m$). Figure 5a) and 5b) show the average CAM maps of Class 6 and Class 2 space-groups. Figure 5c) shows the CAM of an individual, correctly-classified XRD pattern. If we compare the individual CAM to Class 6 CAM, we can see that the neural network is identifying the same reference pattern as most of Class 6 samples, which translates into classification as Class 6. In contrast, Figure 5d) shows an incorrectly classified XRD pattern, which was determined to belong to Class 6 by the CNN, when in reality it belongs to Class 2. A comparison of the individual CAM to the average CAMs of classes 6 and 2, show that the misclassified CAM is more similar to Figure 5a) than Figure 5b). A closer look to the misclassified patterns, shows that the periodicity of peaks before 30° and the relative lack of peaks between 30°-50° (likely caused by mixed phases), are causing the misclassification.

Several representative misclassification cases are identified and included in the Section VII of the Supplementary Information. In our work, by comparing the CAM of wrongly classified cases to the average CAM of all training data within certain class, we conclude there are three main causes for failed classification. The first cause is the mixture of phases in the sample, which increases the chance of the CAM of a particular pattern differing from the average class CAM. The second cause is lack of XRD patterns in the per-class training data. We only have less than 5 patterns in certain space group classes, thus the average CAM of minority classes has only a limited number of discriminative features compared to majority classes, increasing the testing error. The third cause of misclassification is missing peaks, or too few peaks, present in the XRD pattern; becauseof the lack of discriminative features, the CAM of the incorrectly labelled pattern could be ambiguously similar to two or more average class CAMs, causing the misclassification.

After the experimentalist identifies these root-causes of error, they can be mitigated by increasing the number of experimental training points for certain classes or increasing the phase purity of the material coming from the synthesis and solution processing procedure.

**Summary of contributions:**

In this work, we develop a supervised machine learning framework to screen novel materials based on the analysis of their XRD spectra. The framework is designed specifically for cases when only sparse datasets are available, *e.g.*, early-stage high-throughput material development and discovery loops. Specifically, we propose a physics-informed data augmentation method that extends small, targeted experimental and simulated datasets, and captures the possible differences between simulated XRD powder patterns and experimental thin-film XRD patterns. A few thousand augmented spectra are found to increase our classification accuracy from <60% to 93% for dimensionality and 89% for space-group. The $F_1$ *macro* score is also over 0.85 for various algorithms, reflecting the model's capacity to deal with significantly imbalanced classes.

When trained with both augmented simulated and experimental XRD spectra, all convolutional neural networks are found to have the highest accuracy / $F_1$ *macro* score among the many supervised machine learning methods studied. Our proposed a-CNN architecture allows high performance and interpretably through class activation maps. The use of our proposed average class activation maps (average CAMs), allow to identify the root-cause of misclassification, and allow the design of a robust experiment. Furthermore, we find that the neural network model tolerates coarsening of the training data, providing future opportunities for online learning, *i.e.*, the on-the-fly adaptive adjustment of XRD measurement parameters by taking feedback from machine learning algorithms [15].

Our approach can be extended to XRD-based high throughput screening of any kind of thin-film materials, beyond perovskite and perovskite-inspired materials. Since most material databases lack information for various kinds of novel materials, the data augmentation approach to tackle data scarcity can be broadly applied. The underlying difference between XRD patterns of thin-films and powders is common to a broad range of materials and is commonly seen in most thin-film characterization experiments. The high interpretability of our approach could allow future work in semi-supervised or active learning, allowing the CAM maps to guide manual XRD refinement or actual XRD experiments. The framework may also be extended beyond XRD classification, to any spectrum containing information-rich features that require classification (XPS, XRF, PL, mass spectroscopy, etc.). The advantages of our approach include: (1) interpretable error analysis; (2) a data-augmentation strategy that enables fast and accurate classification, even with small and imbalanced data sets.

## V. Methods

### Measurement and pre-processing of experimental XRD patterns

The XRD patterns for each sample are obtained by using a parallel beam, X-ray powder diffraction Rigaku SmartLab system [54] with *2θ* angle from 5-60° with a step size of 0.04°. The tool is configured in a symmetric setup. We preprocess the raw XRD patterns to reduce the experimental noise and the background signal. For this purpose, the background signal is estimated and subtracted along the *2θ* axis, and the spectrum is smoothed conserving the peak width and relative peak size applying the Savitzky-Golay filter [55]. A representative example of a preprocessed and raw XRD spectra is included in Figure S2 in the Supplementary Information.

### Simulation of powder XRD patterns

The powder XRD simulations are carried out with Panalytical Highscore v4.7 software based on the Rietveld algorithm implementation by Hill and Howard [56,57]. The XRD crystal is assumed to have fully random grain orientations. The unit cell lattice parameters, atomic coordinates, atomic displacement parameters, and space group information are considered for the structure factor calculation in the Rietveld model.

### Physics-informed data augmentation:

Suppose we describe the series of peaks in an XRD pattern by a discrete function $(2\theta): I \to \mathbb{R}^+$, which maps a set of discrete angles $I$ to positive real numbers $\mathbb{R}^+$ corresponding to peak intensities. We augment the data through the following sequential process of transformations $\Phi^1_{S,c}, \Phi^2_S$ and $\Phi^3_\varepsilon$:

1. Random peak scaling is applied periodically along the $2\theta$ axis to account for different thin-film preferred orientations. A subset of random peaks at periodic angles $S$ is scaled by factor $c$, such that

$$\Phi^1_{S,c} = c \cdot f|_S + f|_{I\setminus S} \qquad \text{[Eq. 1]}.$$

2. Random peak elimination (with a different randomly-selected $S$ than Eq. 1) is applied periodically along the $2\theta$ axis, to account for different thin-film preferred orientations, such that

$$\Phi^2_S = 0 \cdot f|_S + f|_{I\setminus S} \qquad \text{[Eq. 2]}.$$

3. Pattern shifting by small random value $\varepsilon$ along the $2\theta$ direction to allow for different material compositions and film strain conditions, such that

$$\Phi^3_\varepsilon = f(2\theta - \varepsilon) \qquad \text{[Eq. 3]}.$$

**Class activation maps (CAMs):**

We perform global average pooling in the last convolutional layer. The pooling results are defined as $\sum_x f_k(x)$ where $f_k(x)$ is the activation of unit $k$ at convolution location $x$. The pooling results $\sum_x f_k(x)$ are subsequently fed into the *softmax* classifier. The class confidence score, thus can be computed as:

$$Sc = \sum_k w_k^C \sum_x f_k(x) = \sum_x \sum_k w_k^C f_k(x) \qquad [\text{Eq. 4}].$$

We define $M_C(x) = \sum_k w_k^C f_k(x)$, as the class activation map for class $C$, where $w_k^C$ is the weight of unit $k$ of the last convolution layer of class C. Therefore, $Sc$ can be rewritten as:

$$Sc = \sum_x M_C(x) \qquad [\text{Eq. 5}].$$

$M_C(x)$ directly indicates the importance of the activation at the discrete location $x$ for a given XRD pattern in class $C$.

In the context of XRD and other spectral measurements, the location of relevant intensity peaks is likely to be displaced only little from sample to sample within a certain class. Thus, the class activation maps for all samples within a class are similar, and could be averaged to obtain the discriminative features during activation. Thus, we can define *average class activation maps* for class C as $\overline{M}_C(x)$:

$$\overline{M}_C(x) = \frac{1}{n}\sum_{i=0\,\in C}^{n} M_{C_i}(x) \qquad [\text{Eq. 6}].$$

,where $n$ corresponds to the training patterns labeled as class C.

**Data availability**

The experimental dataset analyzed during the current study is available in the following GitHub repository: [https://github.com/PV-Lab/AUTO-XRD/tree/master/Datasets/Experimental].

The simulated and labelled data that support the findings of this study is available from the Inorganic Crystalline Structure Database (ICSD), but restrictions apply to the availability of these data, which were used under license for the current study, and so are not publicly available. Data are however available from the authors upon reasonable request and with permission of ICSD.

The code used for pre-processing, data augmentation and classification is available at [https://github.com/PV-Lab/AUTO-XRD/]. The classification algorithms are implemented using *scikit-learn* 0.20 [46] and the *fastdtw* python library. The a-CNN is implemented using Keras with the Tensorflow background.

**Acknowledgements and Funding**:

FengXia Wei (A*STAR), Jason Hattrick-Simpers (NIST), Max Hutchinson (Citrine Informatics), and Zachary T. Trautt (NIST) for helpful suggestions on our analysis. Prof. Juan Pablo Correa Baena at Georgia Tech for sharing various experimental XRD patterns. This work was supported by a TOTAL SA research grant funded through MITei (supporting the experimental XRD), the National Research Foundation (NRF), Singapore through the Singapore Massachusetts Institute of Technology (MIT) Alliance for Research and Technology's Low Energy Electronic Systems research program (supporting the machine learning algorithm development), the Accelerated Materials Development for Manufacturing Program at A*STAR via the AME Programmatic Fund by the Agency for Science, Technology and Research under Grant No. A1898b0043 (for ML algorithm error analysis), and by the U.S. Department of Energy under the Photovoltaic Research and Development program under Award DE-EE0007535 (for code framework development). This work made use of the CMSE at MIT, which is supported by NSF award DMR-0819762.

**Competing Interests:**

The Authors declare no Competing Financial or Non-Financial Interests.

**Author Contributions**

FO and ZR contributed equally for this work. FO, ZR, SS, and TB conceived of the research. CS, FO, and SS prepared the simulated XRD data. ZR, FO, CS, BD, and AGK developed data augmentation. SS, NTPH, and CS collected the XRD data. FO, ZR, and GR developed and tested ML algorithms, with key intellectual contributions from RS, ZL, BDC, and AGK. FO and ZR implemented and tuned the final a-CNN architecture and adapted class activation maps to the context of the problem. FO, ZR, SIPT, and TB wrote the manuscript, with input from all co-authors.

**Additional Information**

Supplementary information accompanies the paper on the npj Computational Materials website.

**Figure Legends:**

**Figure 1**. **General Framework and All Convolutional Neural Network Architecture:** a) Schematic of our X-ray diffraction data classification framework, with physics-informed data augmentation. b) Schematic of the best performing algorithm, our All Convolutional Neural Network.

**Figure 2**. **Physics-informed Data Augmentation Algorithm:** Schematic of the physics-informed data augmentation strategy, which accounts for the particularities of thin-film XRD spectra, as described in Eqs. 1–3.

**Figure 3. Accuracy of All Convolutional Neural Network with Augmented Data:** Line plot showing mean Case 3 a-CNN accuracy, as a function of the number of augmented spectra (Methods, Eqs. 1–3) included in the training set, for a) dimensionality classification and b) space-group classification. The *x*-axis shows augmented experimental data (based on the original experimental XRD patterns), and the legend shows simulated data (based on the 164 simulated powder-diffraction patterns obtained from the ICSD). The error bars correspond to one standard deviation from the mean.

**Figure 4. Reducing XRD Pattern Acquisition Time:** Simulation of the trade-off between XRD pattern acquisition rate and predictive accuracy. Accuracies for crystal dimensionality and space

group predictions are estimated by coarsening the XRD spectrum 2*θ* step size for Case 3 conditions. The error bars correspond to one standard deviation from the mean.

**Figure 5. Class Activation Maps For Misclassification Interpretability:** Class activation maps (CAMs) generated by the a-CNN architecture, representing space-group classification of Class 2 ($P2_1/a$) and Class 6 ($Pm\bar{3}m$). Figures a) and b) correspond to maps generated by averaging all training samples in a certain class, while Figures c) and d) corresponds to the CAMs of correctly classified and incorrectly classified individual patterns, respectively. The correctly predicted pattern of c) is explained by the similarity of its CAM to the average CAM of Class 6; whereas, the incorrectly prediction d) can be explained by comparing its CAM to the average CAMs of classes 2 and 6.

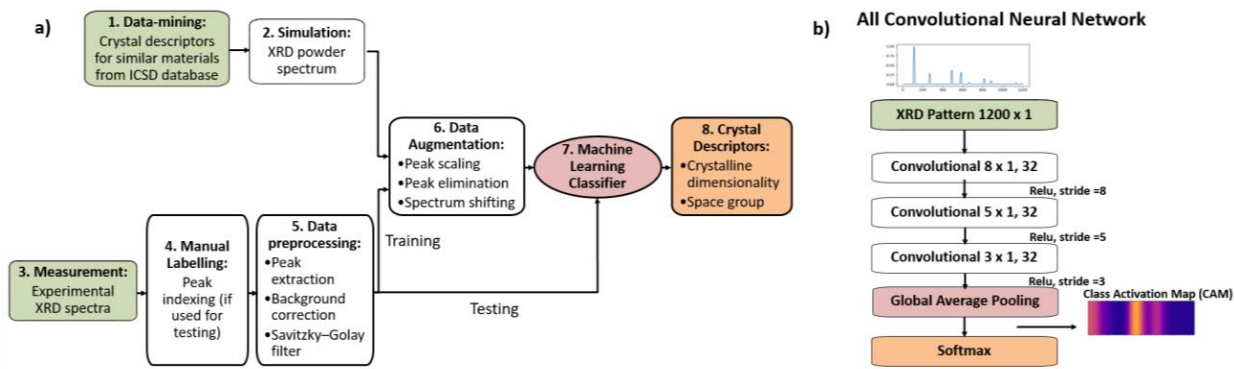

**Figure 1**. a) Schematic of our X-ray diffraction data classification framework, with physics-informed data augmentation. b) Schematic of the best performing algorithm, our All Convolutional Neural Network.

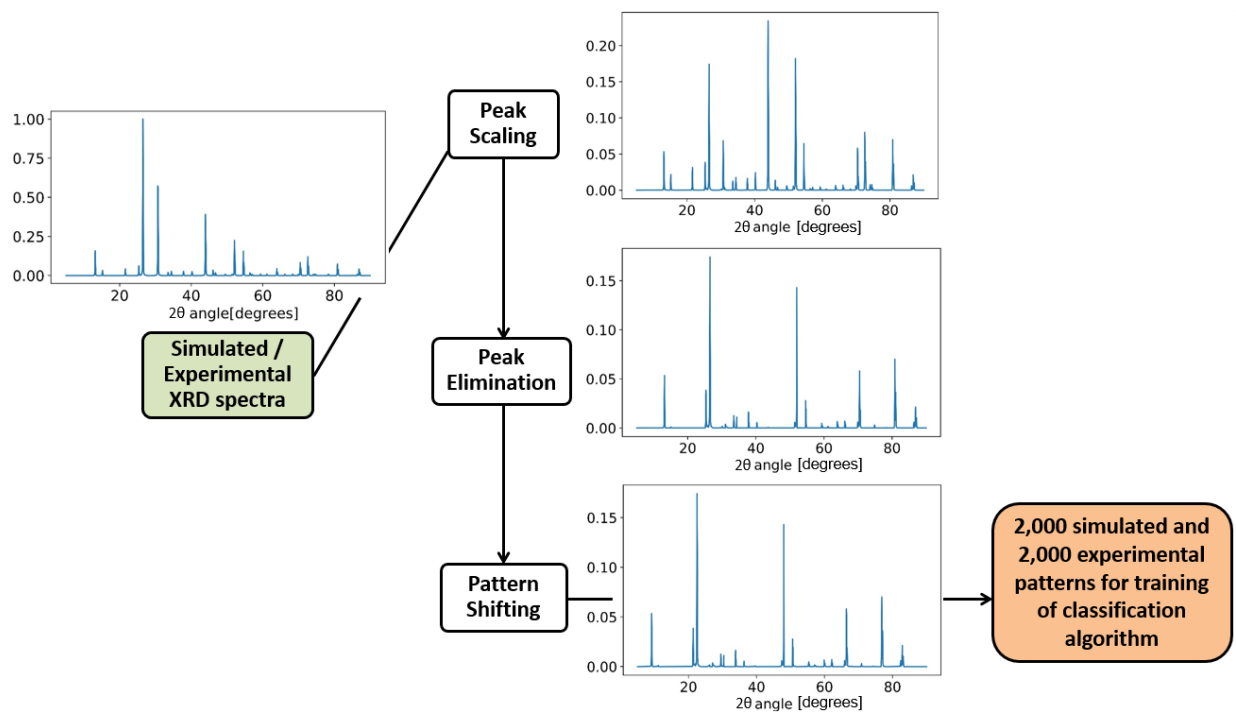

**Figure 2**. Schematic of the physics-informed data augmentation strategy accounts for the particularities of thin-film XRD spectra, as described in Eqs. 1–3.

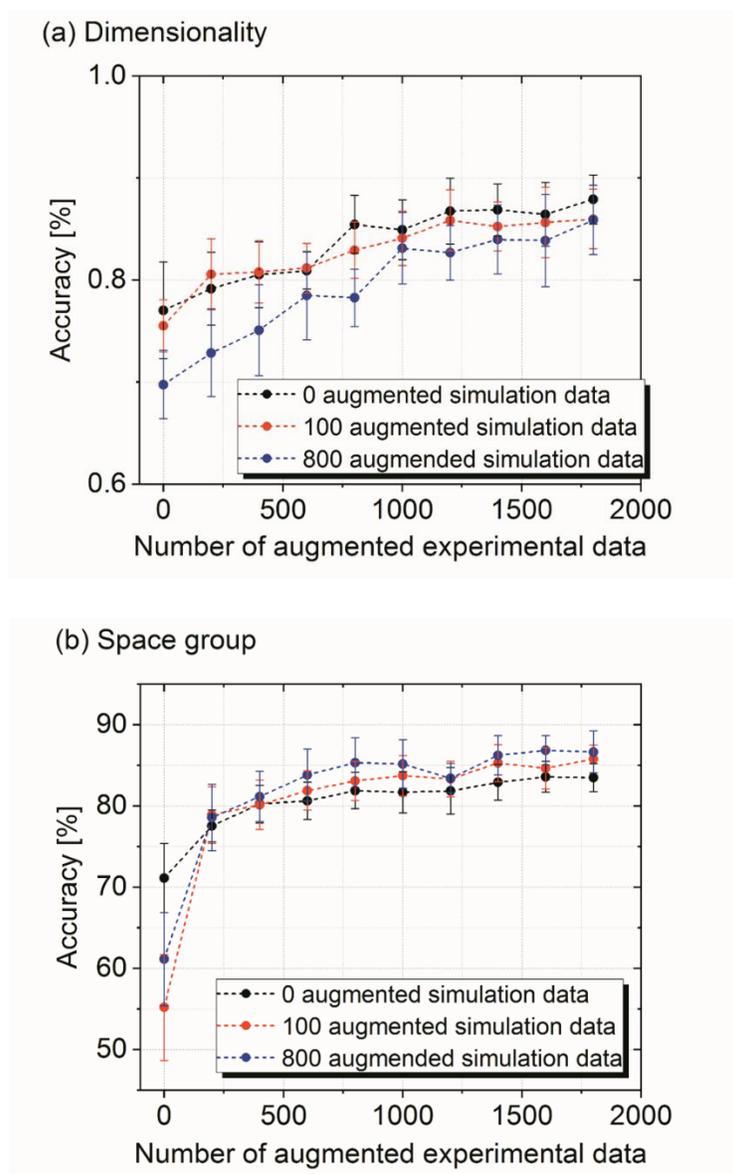

**Figure 3:** Line plot showing mean Case 3 a-CNN accuracy, as a function of the number of augmented spectra (Methods, Eqs. 1–3) included in the training set. The *x*-axis shows augmented experimental data (based on the original experimental XRD patterns), and the legend shows simulated data (based on the 164 simulated powder-diffraction patterns obtained from the ICSD). The error bars correspond to one standard deviation from the mean.

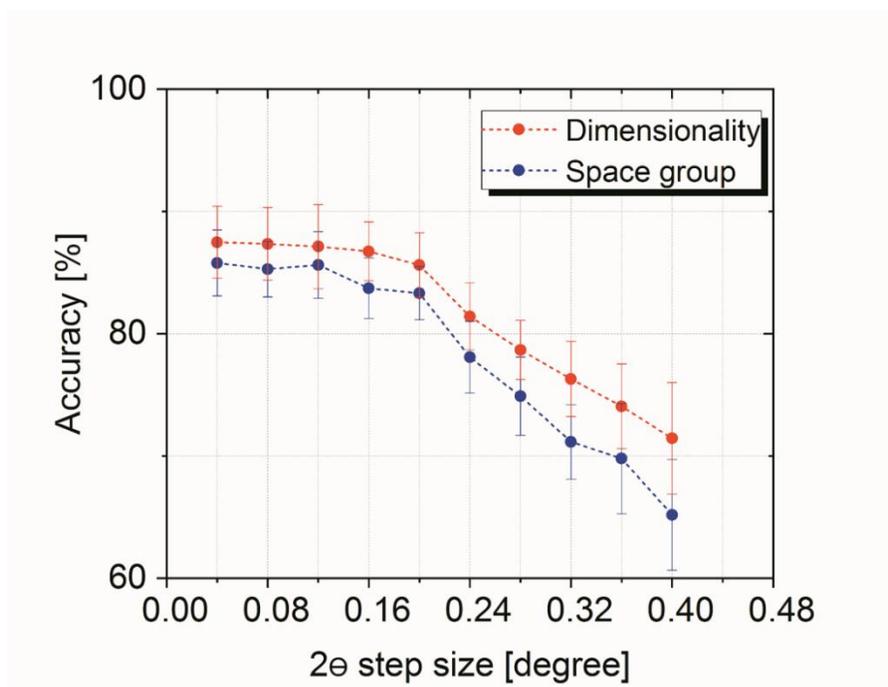

**Figure 4:** Simulation of the trade-off between XRD spectrum acquisition time and a-CNN prediction accuracy. Accuracies for crystal dimensionality and space group predictions are estimated by coarsening the XRD spectrum 2$\theta$ step size for Case 3 conditions. The error bars correspond to one standard deviation from the mean.

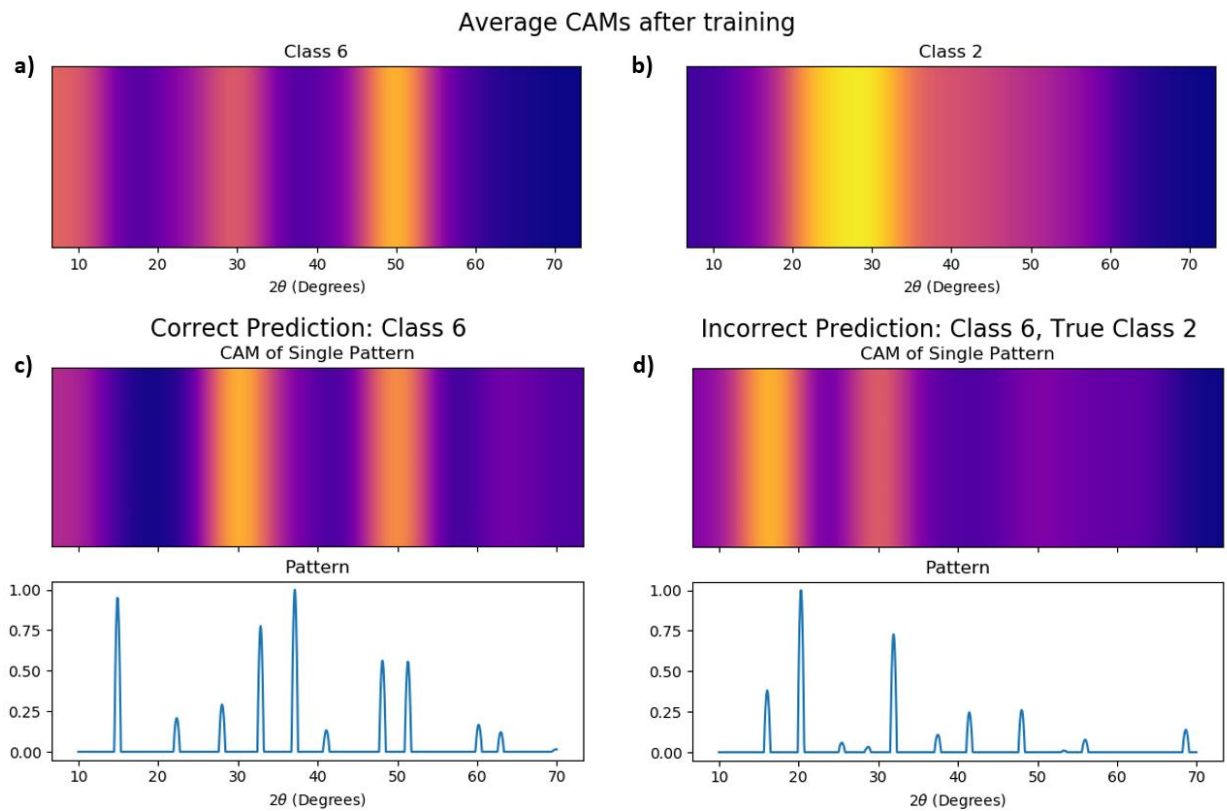

**Figure 5:** Class activation maps (CAMs) generated by the a-CNN architecture, representing space-group classification of Class 2 (P2$_1$/a) and Class 6 ($Pm\bar{3}m$). Figures a) and b) correspond to maps generated by averaging all training samples in a certain class, while Figures c) and d) corresponds to the CAMs of correctly classified and incorrectly classified individual patterns, respectively. The correctly predicted pattern of c) is explained by the similarity of its CAM to the average CAM of Class 6; whereas, the incorrectly prediction d) can be explained by comparing its CAM to the average CAMs of classes 2 and 6.